\documentclass[12pt]{article}
\usepackage{amsmath,amssymb,graphicx}

\newcommand{\point}{\hspace{-17pt}.\hspace{5pt}}

\title{\bf CAN THE GAME \\ BE QUANTUM?}
\author{Andrey
Grib\footnote{e-mail: grib@friedman.usr.lgu.spb.su}~,\quad
Georges Parfionov\footnote{e-mail: your@GP5574.spb.edu} \\
{\small\em Alexandre Friedmann Laboratory of Theoretical
Physics} \\ {\scriptsize St.Petersburg University of
Economics and Finances} \\ {\scriptsize  St.Petersburg,
Russia 191023} \\ {\scriptsize  Fax: +7(812)110-5742,\;
Telephone: +7(812)110-5605}}
\date{}

\begin{document}
\maketitle
\begin{abstract}\baselineskip=5mm
\noindent The game in which acts  of participants don't
have an adequate description in terms of  Boolean logic and
classical theory of probabilities is considered. The model
of the game interaction is constructed on the basis of a
non-distributive orthocomplemented lattice.  Mixed
strategies of the participants are calculated by the use of
probability amplitudes according to the rules of quantum
mechanics. A scheme of quantization of the payoff function
is proposed and an algorithm for the search of Nash
equilibrium is given. It is shown that differently from the
classical case in the quantum situation a discrete set of
equilibrium is possible.
\end{abstract}

\section*{Introduction}
It often occurs that mathematical structures discovered
when solving some class of  problems find their natural
application in totally different areas. The mathematical
formalism of quantum mechanics operating with such notions
as "observable", "state", "probability amplitude" is not an
exception to this rule. The goal of the present paper is to
show that the language of quantum mechanics, initially
applied to the description of the microworld, is adequate
for the description of some macroscopic systems and
situations where Planck's constant plays no role. It is
natural to look for applications of the formalism of
quantum mechanics in those situations when one has
interactions with the element of indeterminacy. So in
recent papers~\cite{Deutsch,Finkelstein,Polley} the
connection of quantum  mechanics with  decision problems in
the conditions of the  indeterminacy is discussed. In
\cite{Gribook} as well as more recently \cite{Waldir} it
was shown that the quantum mechanical formalism can be
applied to description of macroscopical  systems when {\it
the distributive} property for random events is broken. In
the physics of the microworld non-distributivity has an
objective status and must be present in principle. For
macroscopic systems the non-distributivity of random events
expresses some specific case of the observer's "ignorance".

In the present paper a quantum mechanical formalism is
applied to the analysis of a  conflict interaction, the
mathematical model for which is an antagonistic game of two
persons. The game is based on a generalisation of examples
of the macroscopical automata simulating  the behaviour of
some quantum systems considered earlier
in~\cite{GrRZ1,GrRZ2}. A special feature of the game
considered is that the players acts  go in contradiction
with the usual logic. The consequence is breaking of the
classical probability interpretation of the mixed strategy:
the sum of the probabilities for alternate outcomes may be
larger than one. The cause of breaking of the basic
property of the probability is in the {\it
non-distributivity of the logic}. The partners relations
are such that the disjunction "or", conjunction "and" and
the operation of negation do not form  a Boolean algebra
but an orthocomplemented non-distributive lattice. However
this ortholattice happens to be just that which  describes
some properties of a quantum system with spin one half.
This leads to new "quantum" rules for the calculations of
the average profit and new representation of the mixed
strategy, the role of which is played by the "wave
function" -- the normalised vector in a finite dimensional
Hilbert space. Calculations of probabilities are made
according to the standard rules of quantum mechanics.
Differently from the examples of quantum games considered
in~\cite{Eisert,Ekert,Marinatto} where the "quantum" nature
of the game was conditioned by the microparticles or
quantum computers based on them, in our case we deal with a
{\it macroscopic} game, the quantum nature of which has
nothing to do with microparticles. This gives the hope that
our example is one of many analogous situations in biology,
economics etc where the formalism of quantum mechanics can
be used.

%%%%%%%%%%%%%%%%%%%%%%%%%%%%%%%%%%%%%%%%%%%%%%%%%%%%%
\bigskip
\section{\point Where were you Bob?}
\begin{figure}[ht]
\begin{center}
\begin{picture}(100,130)
\linethickness{1pt}

\put(25,115){\circle{15}}

\put(80,115){\vector(-1,0){30}}

\put(0,110){\bf 1} \put(100,110){\bf 2} \put(100,15){\bf 3}
\put(0,15){\bf 4}

\linethickness{2pt}

\put(-16,130){\line(1,0){132}} \put(-16,0){\line(1,0){132}}
\put(-15,0){\line(0,1){130}} \put(115,0){\line(0,1){130}}
\end{picture}
\end{center}
\vspace{-15pt}\caption{\em Bob's ball moves into the place
asked by Alice} \label{fig1}
\end{figure}
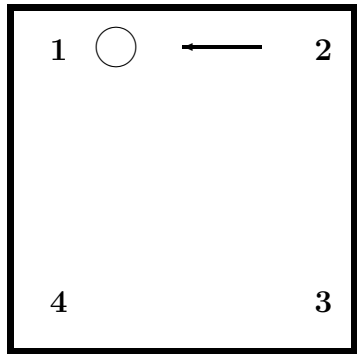

The game "Wise Alice" formulated in our paper is a
modification of the well known game when each of the
participants names one of some previously considered
objects. In the case if the results differ, one of the
players wins from the other some agreed sum of money. The
participants of our game A and B, call them Alice and Bob
have a quadratic box in which a ball is located. Bob puts
his ball in one of the corners of the box but doesn't tell
his partner which corner. Alice must guess in which corner
Bob has put his ball. The rules of the game are such that
Alice can ask Bob questions supposing the two-valued
answer: "yes" or "no". It is supposed that Bob is honest
and always tells the truth. In the case of a "yes" answer
Alice is satisfied, in the opposite case she asks Bob to
pay her some compensation. However, differently from other
such games~\cite{Moulin} the rules of this game (see
Fig.~\ref{fig1}) have one specific feature: {\it Bob has
the possibility to move the ball to any of the adjacent
vertices of the square after
 Alice asks her question.} This additional condition
decisively changes the behaviour of Bob, making him to
become active under the influence of Alice's questions. Due
to the fact that negative answers are not profitable for
him he, in all possible cases, moves his ball to the
convenient adjacent vertex.

%%%%%%%%%%%%%%%%%%%%%%%%%%%%%%%%%%%%%%%%%%%%%%%%%%%%%

%%%%%%%%%%%%%%%%%%%%%%%%%%%%%%%%%%%%%%%%%%%%%%%%%%%%%
\noindent So being in vertices 2 or 4 and getting from
Alice the question "Are you in the vertex~1?" Bob quickly
puts his ball in the asked vertex and honestly answers
"yes". However, if the Bob's ball was initially in the
vertex~3 he cannot escape the negative answer
notwithstanding to what vertex he moves his ball and he
fails. One must pay attention that in this case Alice not
only gets the profit but also obtains the {\it exact
information} on the initial position of the ball: Bob's
honest answer immediately reveals his initial position.

\bigskip
\section{\point Equilibrium-it is when everybody is satisfied!}

The interaction of our players can be described by a four
on four matrix $~(h_{ik})~$  representing payoffs of Alice
in each of the 16 possible game situations
\begin{table}[h]
\begin{center}
\begin{tabular}{|c||c|c|c|c|}
\hline $A \backslash B$ &   {\bf 1} & {\bf 2} & {\bf 3} &
 {\bf 4} \\ \hline\hline
 {\bf ~1~} & ~0~ &  ~0~ & ~a~ & ~0~ \\ \hline
 {\bf ~2~} & ~0~ &  ~0~ & ~0~ & ~b~ \\ \hline
 {\bf ~3~} & ~c~ &  ~0~ & ~0~ & ~0~ \\ \hline
 {\bf ~4~} & ~0~ &  ~d~ & ~0~ & ~0~ \\ \hline
 \end{tabular}
\end{center}\caption{\em The Payoff-matrix of Alice}\label{h1}
\end{table}
\noindent where $a,b,c,d>0$ are her payoffs in those
situations when Bob cannot answer her questions
affirmatively. Our game is an antagonistic game, so the
payoff matrix of Bob is the opposite to that of Alice:
$(-h_{ik})$. The main problem of  game theory is to find
so-called {\it points of equilibrium or saddle points} --
game situations, optimal for all players at once. The
strategies forming the equilibrium situation are optimal in
the sense that they provide to each participant the maximum
of what he/she can get independently of the acts of the
other partner. More or less rational behaviour is possible
only if there are points of equilibrium defined by the
structure of the payoff matrix. A simple criterion for the
existence of the equilibrium points is  known: the payoff
matrix must have the element maximal in its column and at
the same time minimal in its row. It is easy to see that
our game does not have such equilibrium points.
Non-existence of the saddle point follows from the strict
inequality valid for our game $$\max_{j}\min_{k}h_{jk}<
\min_{k}\max_{j}h_{jk}$$ So there are no stable strategies
to follow for Bob and Alice in each {\it separate} turn of
the game. In spite of the absence of a rational choice at
each turn of the game, when the game is repeated many times
some optimal lines of behaviour can be found. To find them
one must, following von Neumann~\cite{Neumann}, look for
the so called mixed generalisation of the game. In this
generalised game the choice is made between {\it mixed
strategies} i.e. probability distributions of usual (they
are called differently from mixed "pure" strategies)
strategies. As the criterion for the choice of optimal
mixed strategies one takes the mathematical expectation
value of the payoff which shows how much one can win on
average by repeating the game many times. The optimal mixed
strategies for Alice and Bob are defined as such
probability distributions on the sets of pure strategies
$x^0=(x^0_1,x^0_2,x^0_3,x^0_4)$ and
$y^0=(y^0_1,y^0_2,y^0_3,y^0_4)$ that for all distributions
of $x, y$ the von Neumann-Nash inequalities are valid:
\begin{equation}\label{NE}
       {\cal H}_A (x^0,y^0)\geqslant {\cal H}_A (x,y^0)\,,\qquad {\cal H}_B
       (x^0,y^0)\geqslant {\cal H}_B (x^0,y),
\end{equation}
where { \it$\cal{H}_{A},\cal{H}_{B}$ -- payoff functions}
of Alice and Bob are the expectation values of their wins
$${\cal H}_A(x,y) = \sum_{j,k=1}^{4}{h_{jk} x_j y_k}\,,
\qquad {\cal H}_B(x,y) = -\sum_{j,k=1}^{4}{h_{jk}x_j y_k}$$
The combination of strategies, satisfying the von
Neumann-Nash inequalities, is called {\it the situation of
equilibrium} in Nash's sense. The equilibrium is convenient
for each player, deviation from it  can only make the
profit smaller. In equilibrium situations the strategy of
each player is optimal against the strategy of his (her)
partner. Existence of equilibrium in mixed strategies is
based on the main theorem of matrix games theory (von
Neumann's theorem). To find them one must solve the pair of
dual problems of linear programming and it is made easily.
The only question is: do optimal strategies correctly
describe the behaviour of Bob and Alice in their game with
a ball?

\medskip
\section{\point The classical "Foolish Alice"}

In classical matrix game theory the optimal strategies of
the players are totally defined by their interests. All
other characteristics of the participants of the game are
totally ignored. To go from this oversimplification of von
Neumann's game theory one must look for other concepts of
equilibrium, for example due to von
Stackelberg~\cite{Moulin} or to study influence of the
psychological relation on the outcomes of
games~\cite{Kemeny}. Our attention will be concentrated not
on the psychological but on the {\it logical} aspect of the
conflict interaction of players. Before discussing the
logical nuances pay attention to the fact that the payoff
matrix in table~(\ref{h1}) does not give full information
about {\it the rules} of the game and interactions of the
players.

To see this consider the totally different (from the point
of view of the behaviour of players the antagonistic) game
"The foolish Alice" with the same payoff matrix as the game
with a ball. Alice and Bob decide to meet at the corner of
a big four-corner house but don't agree on which corner. As
usual Bob comes first. If Alice comes to a corner from
which she can see Bob she is satisfied, in the opposite
case she, thinking that he didn't come, retires being
insulted.
 The next day Bob, in order to calm her, must give her some
expensive present. Differently from the previous game each
participant of this game has {\em passive} position. If Bob
does not see Alice he has no reasons to go from one corner
of the building to the other because he does not know if
she came or is just standing in the opposite corner. How to
discriminate these two identical (in the structure of the
payoff) games?

%%%%%%%%%%%%%%%%%%%%%%%%%%%%%%%%%%%%%%%%%%%%%%%%%%%%%
\begin{figure}[ht]
\begin{center}
\includegraphics[height=0.45\textwidth]{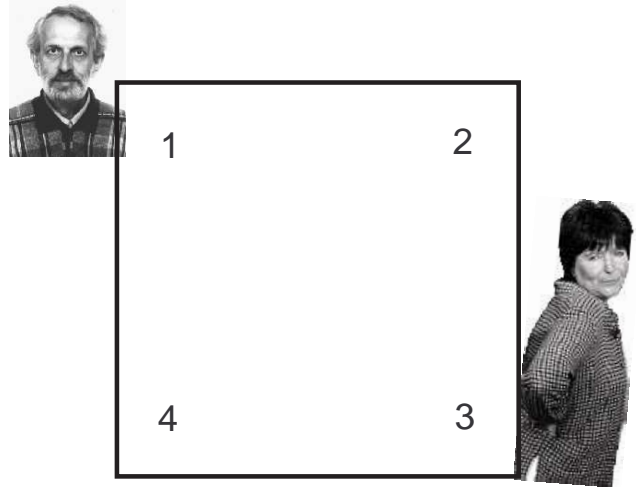}
\end{center}
\vspace{-20pt}\caption{\em Alice does not see Bob and is
very dissatisfied} \label{fig2}
\end{figure}

\noindent In order to see clearly the difference between
the two games and to discriminate "wise Alice" from the
"foolish" one we introduce notations making the difference
evident. Encode the strategies of Alice and Bob by vectors,
consisting of zeros and ones:
$$\alpha=(\alpha_1,\alpha_2,\alpha_3,\alpha_4),\qquad
\beta=(\beta_1,\beta_2,\beta_3,\beta_4) $$ so that the
component equal to one means the applied pure strategy.
Then it is evident that
$$\sum_{j=1}^4{\alpha_j}=\sum_{j=1}^4{\beta_j}=1,\qquad
\alpha_j\cdot\alpha_k=\beta_j\cdot\beta_k=\delta_{jk}$$ and
the profit of Alice {\it in one turn} of any of the games
considered is
\begin{equation}
\label{Halica} {\cal H}_A(\alpha,\beta) =
\sum_{j,k=1}^{4}{h_{jk}\alpha_j \beta_k}
\end{equation}

\noindent So, in the separate turn the "wise" Alice is not
different from the "foolish" one. The difference occurs in
{\it the behaviour} when the game is repeated many times.
The source of the difference is in the different method of
calculation of {\it the average} payoff.

Consider it explicitly. At first let us take the classical
case of interaction. In the case of the "foolish" Alice the
initial strategies of Bob are not correlated with the
strategies of his partner, so $$E(\alpha_j\cdot
\beta_k)=E\alpha_j \cdot E\beta_k$$ and averaging of the
payoff gives the well known classical expression $$E{\cal
H}_A(\alpha,\beta)= \sum_{j,k=1}^{4}{h_{jk} E\alpha_j\cdot
E\beta_k}=\sum_{j,k=1}^{4}{h_{jk}x_j y_k}={\cal H}_A(x,y),
$$ where $x_j, y_k$ are frequencies of the corresponding
pure strategies. For our payoff matrix one obtains the
expression of the payoff function for Alice as
\begin{equation}\label{Halxy}
{\cal H}_A(x,y)=ax_1 y_3+bx_2 y_4+cx_3 y_1+dx_4 y_2
\end{equation}
Then one must, using the linear programming, find the
saddle points for natural constrains
$$x_1+x_2+x_3+x_4=1,\quad y_1+y_2+y_3+y_4=1,\qquad
x_j,\,y_k \geqslant 0.$$

In our case there is only one equilibrium point and the
mixed strategies of Alice and Bob are found as: $$x=(\mu
a^{-1},\; \mu b^{-1},\; \mu c^{-1},\; \mu d^{-1}),\qquad
y=(\mu c^{-1},\; \mu d^{-1},\; \mu a^{-1},\; \mu b^{-1}),$$
where $\mu= {(a^{-1}+ b^{-1}+ c^{-1}+ d^{-1})}^{-1}$ is the
price or the value of the game i.e. the average profit of
Alice in the equilibrium situation. So the optimal
frequency of Bob's being in this or that corner of the
building is inversely proportional to the sum of money
which he must he give to his girl friend. The optimal
strategy of Alice is more sophisticated: she must not be
very greedy and more frequently come to the places where
her friend will not pay too much to her.

Instead let us write the expression of the payoff function
of the "foolish" Alice in a somehow different form, useful
for our subsequent considerations. Consider random events
$\alpha_{13},\,\alpha_{24}\subset {\cal
S}_A,\;\;\beta_{13},\,\beta_{24}\subset {\cal S}_B,$ where
${\cal S}_A,\,{\cal S}_B =\{1, 2, 3, 4 \}$ -- the sets of
pure strategies of Alice and Bob. Each of the considered
events corresponds to the choice of the pair of opposite
corners of the house. It is easy to see that events
$\alpha_{jk} \cap\beta_{lm}$ form the ${\cal S}_A \times
{\cal S}_B$ division of the space of the game situations
and thus form a complete set of events. Taking this into
account one can write the payoff function for Alice as the
mixture of conditional expectation values. $$ {\cal
H}_A(x,y)=\sum_{jk,lm}{E_{\alpha_{jk}\cap \beta_{lm}}{\cal
H}_A\cdot P(\alpha_{jk}\cap \beta_{lm})} $$ In the case of
our payoff matrix this expression has the form
\begin{equation} \label{classic} (ap_{13}^1
q_{13}^3 + cp_{13}^3 q_{13}^1)\cdot
P(\alpha_{13}\cap\beta_{13}) + (bp_{24}^2 q_{24}^4 +
dp_{24}^4 q_{24}^2)\cdot P(\alpha_{24} \cap \beta_{24}),
\end{equation}
where  $p_{jk}^l,\, q_{jk}^l$ -- conditional probabilities
of the choice of the corner $l$ from the given pair of
opposite corners $\{j, k\}$. From this it follows that the
conditional average payoff for Alice will be
\begin{equation}\label{Hal13}
E_{\alpha_{13}\cap \beta_{13}}{\cal H}_A = ap_{13}^1
q_{13}^3 + cp_{13}^3 q_{13}^1
\end{equation}
if both players choose the diagonal $\{1,3\}$ and
\begin{equation}\label{Hal24}
E_{\alpha_{24}\cap \beta_{24}}{\cal H}_A =bp_{24}^2
q_{24}^4 + dp_{24}^4 q_{24}^2
\end{equation}
if they prefer the diagonal $\{2,4\}$. If the players
choose different diagonals of the house the conditional
payoff is equal to zero because in this case it is always
possible for Alice to see Bob and he must not pay for
presents. Easy calculations show that
$$P(\alpha_{jk}\cap\beta_{jk})=(x_j+x_k)(y_j+y_k),\quad
p_{jk}^l=\frac{x_l}{x_j+x_k},\quad
q_{jk}^l=\frac{y_l}{y_j+y_k} $$ These formulas will be of
use for us when we discuss the behaviour of Bob moving the
ball in the game "Wise Alice".
%%%%%%%%%%%%%%%%%%%%%%%%%%%%%%%%%%%%%%%%%%%%%

\bigskip
\section{\point Different logics -- different behaviour}

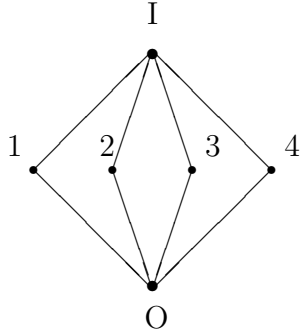
\begin{figure}[ht]
\begin{center}
\begin{picture}(120,120)
\put(45,94){\circle*{4}} \put(43,105){I}
\put(0,50){\circle*{3}} \put(-10,55){1}
\put(30,50){\circle*{3}} \put(25,55){2}
\put(60,50){\circle*{3}} \put(65,55){3}
\put(90,50){\circle*{3}} \put(95,55){4}
\put(45,6){\circle*{4}} \put(42,-10){O}
\put(0,50){\line(1,1){45}} \put(0,50){\line(1,-1){45}}
\put(90,50){\line(-1,1){45}} \put(90,50){\line(-1,-1){45}}
\put(30,50){\line(1,3){15}} \put(30,50){\line(1,-3){15}}
\put(60,50){\line(-1,3){15}} \put(60,50){\line(-1,-3){15}}
\end{picture}
\end{center}
\vspace{-15pt}\caption{\em Lattice of Alice's questions and
Bob's answers} \label{fig3}
\end{figure}

Let us discuss now the behaviour of players in the game
"Wise Alice". First notice that this game gives the
simplest model of measurement (defining the place of the
object). Alice wants to know where is Bob located but she
makes Bob active by her questions, "preparing" him in a
definite "state". She gets exact information, not in all
cases, but only when the negative answer is obtained. Alice
makes proposals but the logic of her propositions must be
somehow different from the classical scheme. More
explicitly, let  $\alpha_k$ is the proposition of Alice
that Bob's ball is located on vertex number $k$. One can
consider this value as the predicate: the function defined
on the set $S_B$ of {\it initial} strategies of Bob taking
logical values 0 or 1. For our box with a  ball
Fig.~\ref{fig1} it is easy to see that the values of
propositions of  Alice are distributed as follows: $$
\alpha_1(1)=\alpha_1(2)=\alpha_1(4)=1,\quad \alpha_1(3)=0,
$$
\begin{equation}\label{Valent}
\alpha_2(2)=\alpha_2(1)=\alpha_2(3)=1,\quad \alpha_2(4)=0,
\end{equation}
$$\alpha_3(3)=\alpha_3(2)=\alpha_3(4)=1,\quad
\alpha_3(1)=0,$$
$$\alpha_4(4)=\alpha_4(1)=\alpha_4(3)=1,\quad
\alpha_4(2)=0.$$ Defining disjunction as usual as
$$\alpha_j\vee\alpha_k=\max(\alpha_j,\alpha_k)$$ one
obtains a "slight" breaking of the classical logic: the
disjunction of any pair of different propositions occurs to
be identically true: $$\alpha_j\vee\alpha_k=1,\qquad j\neq
k$$ Remember that for the classical "foolish" Alice one has
$$\alpha_1\vee\alpha_2\vee\alpha_3\vee\alpha_4=1$$
Differences with classical logic occur also for negation.
Instead of the classical relations
$\neg\alpha_j=1-\alpha_j$ one has the equalities:
$$\neg\alpha_1=\alpha_3,\quad\neg\alpha_3=\alpha_1,\quad
\neg\alpha_2=\alpha_4,\quad\neg\alpha_4=\alpha_2$$ Really,
every time when Alice learns that the Bob's ball  is {\it
not} located at the questioned vertex of the square she
understands that it is located at the opposite vertice.
Notice that the law of double negation
$\neg(\neg\alpha_j)=\alpha_j$ as well as the law of the
excluded third $\alpha_j\vee(\neg\alpha_j)=1$ are valid.
One must define the conjunction. It can be introduced by
the standard formula
$$\alpha_j\wedge\alpha_k=\neg((\neg\alpha_j)\vee(\neg\alpha_k))$$
It is easy to see that this is the only way of defining the
conjunction if De Morgan's laws of duality are valid:
$$\neg(\alpha_j\wedge\alpha_k)=(\neg\alpha_j)\vee(\neg\alpha_k),\qquad
\neg(\alpha_j\vee\alpha_k)=(\neg\alpha_j)\wedge(\neg\alpha_k)$$
Defining thus all logical operations let us check other
differences with classical Boolean logic. First notice that
in spite of the fact that any pair of different
propositions of Alice is in complementarity:
$$\alpha_j\vee\alpha_k=1,$$ $$\alpha_j\wedge\alpha_k=0,$$
not all of them are {\it orthocomplemented}-- mutually
opposite. Only pairs of propositions with the same "parity"
$\{\alpha_1;\alpha_3\}$ $\{\alpha_2; \alpha_4\}$ are
orthocomplemented. Second, one has a breaking of the
distributivity law. So, for any triple of different $j,k,l$
one has the inequality
$$(\alpha_j\vee\alpha_k)\wedge\alpha_l \neq
(\alpha_j\wedge\alpha_l)\vee(\alpha_k\wedge\alpha_l)$$
Really, the left side of the inequality is equal to
$\alpha_l$,while the right side is zero. So the logic of
Alice occurs to be {\it a non-distributive
orthocomplemented lattice}~\cite{Birkhoff,Gratzer}. Same
concerns the logic of Bob. His state reduced by the
question of Alice is described by the analogous system of
predicates $\beta_1,\beta_2, \beta_3, \beta_4$ defined on
the set of strategies of Alice and taking the value 1 for
those questions on which he can answer affirmatively. In
Fig.~\ref{fig3} a Hasse diagram is shown describing the
nondistributive logic of our players. In spite of the
logical differences of introduced predicates $\alpha_1,
\alpha_2, \alpha_3, \alpha_4$ and $\beta_1, \beta_2,
\beta_3, \beta_4$ from the classical ones one can think
that the payoff of Alice in the {\it separate turn} of the
game still is given by the same expression as before
$${\cal H}_A(\alpha,\beta) =
\sum_{j,k=1}^{4}{h_{jk}\alpha_j\beta_k}$$ The main
difficulties arise when one goes to the repeated game and
when one tries to calculate the average payoff.  If one
tries to give to the average value of predicates the
probabilistic interpretation: $$E\alpha_j=x_j,\qquad
E\beta_k=y_k,$$ one immediately comes to the contradiction:
the sum of probabilities of pair-wise disjoint (due to our
definition of the conjunction) outcomes is larger than
unity $$x_1+x_2+x_3+x_4=3,\qquad y_1+y_2+y_3+y_4=3.$$ This
follows from the additivity of the average and (8), leading
to the following identities
$$\alpha_1+\alpha_2+\alpha_3+\alpha_4=3,\qquad
\beta_1+\beta_2+\beta_3+\beta_4=3$$ Hope for the validity
of additivity for pairs of mutually neglecting events also
occurs to be in vain. Due to the property that the
left-handed side of easily checked inequalities
\begin{equation}\label{bad}
\alpha_1+\neg(\alpha_1)\geqslant I,\quad
\alpha_2+\neg(\alpha_2)\geqslant I,\quad
\alpha_3+\neg(\alpha_3)\geqslant I,\quad
\alpha_4+\neg(\alpha_4)\geqslant I
\end{equation}
sometimes takes the value equal to 2, the main property of
 probability is broken even for orthocomplemented
elements. The probability properties are also broken for
the disjunction (see Hasse diagram Fig.~\ref{fig3}).

%%%%%%%%%%%%%%%%%%%%%%%%%%%%%%%%%%%%%%%%%%%%%%%%%%%%%
%%%%%%%%%%%%%%%%%%%%%%%%%%%%%%%%%%%%%%%%%%%%%%%%%%%%%

\noindent If one considers all outcomes equally possible,
then the probability of the always true event, i.e.
disjunction of any of two events occurs to be one half! So
a classical probabilistic description of the behaviour of
the players in the repeated game is impossible in
principle. The solution for the situation arising is given
by the ideas of quantum mechanics.
%%%%%%%%%%%%%%%%%%%%%%%%%%%%%%%%%%%%%%%%%%%%%%%%%%%%%%%%%%%%

\medskip
\section{\point To averages through quantization}

Following A.A.Grib and R.R.Zapatrin{\cite {GrRZ1} we pay
attention to the fact that the ortholattice of the logic of
interaction of partners of the "Wise Alice" is isomorphic
to the ortholattice of invariant subspaces of the Hilbert
space of the quantum system with spin $\frac{1}{2}$ and
observables of the type of $S_x$ $S_\theta$.
%%%%%%%%%%%%%%%%%%%%%%%%%%%%%%%%%%%%%%%%%%%%%%%%%%%%%
\begin{figure}[ht]
\begin{center}
\begin{picture}(200,160)

\qbezier[60](120,90)(119,115)(105,120) \put(80,100){$0$}
\put(125,100){\Large $\theta$}

\linethickness{1pt}
\put(90,90){\line(1,0){90}}\put(90,90){\line(-1,0){90}}\put(170,95){$a1$}
\put(90,90){\line(0,1){70}}\put(90,90){\line(0,-1){70}}\put(75,150){$a3$}
\put(90,90){\line(1,2){37}}\put(90,90){\line(-1,-2){35}}\put(45,30){$a2$}
\put(90,90){\line(2,-1){90}}\put(90,90){\line(-2,1){90}}\put(190,35){$a4$}
\end{picture}
\end{center}
\vspace{-20pt}\caption{\em Lattice of invariant subspaces
of observables $S_x$ and $S_\theta$} \label{fig4}
\end{figure}
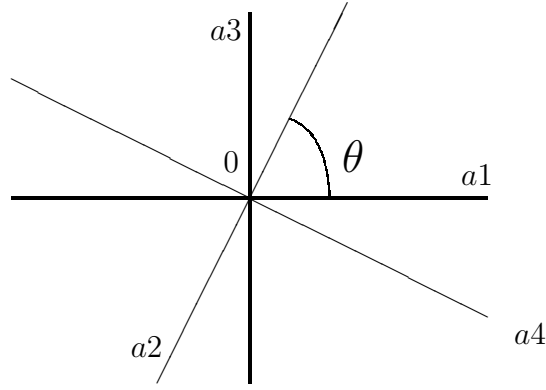
%%%%%%%%%%%%%%%%%%%%%%%%%%%%%%%%%%%%%%%%%%%%%%%%%%%%%
In Fig.~\ref{fig4} two pairs of mutually orthogonal direct
lines $\{a1; a3\}$, $\{a2; a4\}$ are shown. One of these
pairs makes diagonal the operator $S_x$, the other
$S_\theta$. If one takes as representations of logical
conjunction and disjunction their intersection and linear
envelope and if negation corresponds to the orthogonal
complement one obtains the ortholattice isomorphic to the
logic of our players. One example of such an isomorphism is
the mapping $\alpha_j \mapsto aj,\; j=1, 2, 3, 4$. We saw
that in one "experiment" neither Alice nor Bob have a
stable strategy. However if the game is repeated many times
one can ask about optimal frequencies of the corresponding
pure strategies. Due to the non-distributivity of the logic
,as we saw previously, it is impossible to define on the
sets $S_A$ and $S_B$ of pure strategies a probabilistic
measure. The main problem is calculation of an adequate
{\it procedure of averaging}.

Following well known constructions of quantum mechanics we
take instead of the sets of pure strategies of Alice and
Bob $S_A,S_B$ the pair of two-dimensional Hilbert spaces
$H_A,H_B$. So {\it pure strategies} are represented by
one-dimensional subspaces or {\it normalised vectors} of
Hilbert space (wave functions). Use of Hilbert space
permits us without any difficulties to realise the
non-distributive logic of our players. For this one must
represent predicates $\alpha_1, \alpha_2, \alpha_3,
\alpha_4,$ ¨ $\beta_1, \beta_2, \beta_3, \beta_4$
describing questions of Alice and locations of Bob's ball
by the corresponding self-conjugate operators. It is
important to notice that the ortholattice can be realised
in an infinite set of ways. Describe all these ways up to
the unitary equivalence. We do this for the predicates of
Alice. The same can be done for Bob. In $H_A$ take an
arbitrary one-dimensional subspace and put its
orth-projector into correspondence to the predicate
$\alpha_1$: the projector denote as $\widehat{\alpha}_1$.
To the predicate $\alpha_3$ put into correspondence {\it
the orthogonal complementary} projector
$\widehat{\alpha}_3$. Then the following evident and for us
desirable relations are valid:
\begin{equation}\label{reduce13}
   \widehat{\alpha}_1\cdot\widehat{\alpha}_3=
   \widehat{\alpha}_3\cdot\widehat{\alpha}_1=O,\qquad
   \widehat{\alpha}_1 +\widehat{\alpha}_3=I
   \end{equation}
It is clear that the choice of these operators is unique up
to unitary equivalence. Take an arbitrary one-dimensional
subspace {\it different from the eigenspaces of the
projector}~$\widehat{\alpha}_1$ and put it into
correspondence to the predicate of some orth-projector
$\widehat{\alpha}_2$. To the variable $\alpha_4$ put the
projector $1-\widehat{\alpha}_2$ . In the result one
obtains one more resolution of Hilbert space $~H_A$
    \begin{equation}\label{reduce24}
\widehat{\alpha}_2\cdot\widehat{\alpha}_4=
\widehat{\alpha}_4\cdot\widehat{\alpha}_2=O,\qquad
\widehat{\alpha}_2 +\widehat{\alpha}_4=I
\end{equation}
Projectors from different resolutions do not commute one
with another. However there is a simple connection between
them: the second of direct resolutions is obtained from the
first by rotation through an angle different from multiples
of $90^o$. In other words there exists a unitary operator
$~u~$ such that
\begin{equation}\label{unitary} \widehat{\alpha}_2 = u^{-1}
\widehat{\alpha}_1 u ,\qquad \mbox{¨}\qquad
\widehat{\alpha}_4 = u^{-1} \widehat{\alpha}_3 u
\end{equation}
It is clear that this unitary operator defines the class of
unitary equivalent realisations of our non-distributive
ortholattice, negation is represented by going to the
complementary projector. Notice that differently from the
predicative description due to rules~(\ref{Valent}) leading
to unpleasant inequalities~(\ref{bad}) the operator
representation of the non-distributive lattice due
to~(\ref{reduce13},\ref{reduce24}) is "friendly" to
orthocomplementarity:
$$\widehat{\alpha}_1+(\widehat{\alpha}_1)^\bot =I,\quad
\widehat{\alpha}_2+(\widehat{\alpha}_2)^\bot =I,\quad
\widehat{\alpha}_3+(\widehat{\alpha}_3)^\bot =I,\quad
\widehat{\alpha}_4+(\widehat{\alpha}_4)^\bot =I$$ Doing the
same for the Bob's lattice one comes to the observables
$\widehat{\beta_1}, \widehat{\beta_2}, \widehat{\beta_3},
\widehat{\beta_4}$ and analogous to $~u~$ a unitary
operator $~v~$ giving the connections between "even" and
"odd" variables. The next step of our scheme is {\it the
space of game situations}. The quantum analog of the
classical space of game situations becomes the tensor
product~${\sf H}_A\otimes{\sf H}_B$. This space is
necessary for introducing the main observable: the payoff
for Alice. To write this observable following quantum
mechanics take the classical  expression~(\ref{Halxy}) of
the payoff function of Alice $${\cal H}_A(\alpha,\beta) =
\sum_{j,k=1}^{4}{h_{jk}\alpha_j \beta_k}$$ and write there
the corresponding projectors. In the result one obtains the
self-conjugate operator in ${\sf H}_A\otimes{\sf H}_B$,
{\it the observable} of the payoff for Alice:
$$\widehat{{\cal H}}_A =\sum_{j,k=1}^{4}h_{jk}
\widehat{\alpha}_j\otimes \widehat{\beta}_k$$ Let Alice and
Bob repeat their game with a ball many times and let us
describe their{\it behaviour} by normalised vectors
$\varphi\in{\sf H}_A, \;\mbox{¨}\; \psi\in{\sf H}_B$. The
element $s=\varphi\otimes\psi$ expressing their interaction
during the game is a normalised vector in ${\sf
H}_A\otimes{\sf H}_B$. Taking it as the characteristic of
the {\it state} of the game calculate the average in this
state $E_{\varphi\otimes\psi}\widehat{{\cal H}}_A$
according to the standard rules of quantum mechanics
$$\langle\widehat{{\cal H}}_A s,\, s\rangle
=\sum_{j,k=1}^{4}{h_{jk} \langle(\widehat{\alpha}_j\otimes
\widehat{\beta}_k)\;\varphi\otimes\psi,\;
\varphi\otimes\psi\rangle }$$ so that after easy
transformations one gets for the average payoff for the
given types of behaviour of the players:
$$E_{\varphi\otimes\psi}\widehat{{\cal H}}_A
=\sum_{j,k=1}^{4}{h_{jk}
\langle\widehat{\alpha}_j\varphi,\, \varphi\rangle \cdot
\langle\widehat{\beta}_k\psi,\, \psi\rangle }$$ Putting
into this formula the elements of our payoff matrix and
using the notations
\mbox{$~p_j=\langle\widehat{\alpha}_j\varphi,\,\varphi\rangle~$},
$~q_k=\langle\widehat{\beta}_k\psi,\,\psi\rangle~$ one
obtains
\begin{equation}\label{Hquant}
E_{\varphi\otimes\psi}\widehat{{\cal H}}_A =ap_1 q_3 +cp_3
q_1 + bp_2 q_4 + dp_4 q_2
\end{equation}
It is useful to compare this expression with the classical
average obtained earlier for the "foolish" Alice: $${\cal
H}_A(x,y)=ax_1 y_3+cx_3 y_1+bx_2 y_4+dx_4 y_2 $$ There is
some resemblance but there is also a serious difference.
One could ask, why it is impossible to recalculate the
quantum average differently by normalising the frequencies
to one not only for each diagonal separately but for two
diagonals as it is made in the classical case? But the fact
is that like in case of measuring non-commuting operators
of spin projections Alice knows about her interaction with
Bob when she asks him and receives the answer. This
interaction is different when questions on different
diagonals are asked because different positions of the ball
are immovable for these cases. So relative to different
interactions (different context) different events are
defined leading to different probabilistic spaces as it is
true for measuring spin projections. One can use the
metaphor that if in one case Alice is throwing the coin and
is interested if "up" or "down" will arise, in the other
case the interaction with Alice will change the coin in
such a way as if a new coin is thrown, on one side of which
a big "up" and the small "down" of the previous  coin are
drawn. On the other side of the new coin the opposite
situation occurs. So the new coin is made asymmetric
following the structure of the payoff matrix. In
homogeneity of events in the quantum case makes impossible
the renormalisation of frequencies and leads to new results
for averages.
%%%%%%%%%%%%%%%%%%%%%%%%%%%%%%%%%%%%%%%%%%%

\bigskip
\section{\point Probability amplitudes instead of probabilities}

The main difference in the given formulas is in the {\it
sense} of variables. If $x_j,x_k$ are probabilities, the
$~p_j,p_k~$ cannot be  because they contradict the main
laws of the probability theory. Really, taking into account
that the projectors $\widehat{\alpha}_j$ and
$\widehat{\beta}_k$ with indices of the same parity commute
and form a resolution of unity, one obtains after standard
calculations the following identities.
  \begin{equation}\label{amplitude} p_1+p_3= \langle
\widehat{\alpha}_1\varphi,\, \varphi\rangle
+\langle\widehat{\alpha}_3\varphi,\, \varphi\rangle
=\langle(\widehat{\alpha}_1+\widehat{\alpha}_3) \,
\varphi,\varphi\rangle=1,
\end{equation}
\begin{equation}
p_2+p_4=1,\quad q_1+q_3=1,\quad q_2+q_4=1
\end{equation}
So, differently from classical probability theory, for each
family of pairwise disjoint events
$\{\widehat{\alpha}_j\},\,\{\widehat{\beta}_k\}$  one
obtains the relations $$ p_1+p_2+p_3+p_4=2,\qquad
q_1+q_2+q_3+q_4=2$$ The sense of the values $~p_j,p_k$ is
obtained by using the standard quantum rule. Let for
example the behaviour of Alice be described by the
normalized vector $\varphi\in{\sf H}_A$, and let  
$\xi_1^+, \xi_1^-\in{\sf H}_A$ be normalised eigenvectors
of the projector with eigenvalues 1 ("yes") and 0 ("no").
Projecting the state vector of Alice onto the basis
$\{\xi_1^+, \xi_1^-\}$ $$\varphi=c_+ \xi_1^+ + c_-
\xi_1^-$$ one obtains that she can find Bob's ball on the
first vertex of the square with the probability $$ p_1
=\langle\widehat{\alpha}_1(c_+ \xi_1^+ + c_- \xi_1^-),\;
c_+ \xi_1^+ + c_- \xi_1^-\rangle =\langle c_+ a_1^+ ,\; c_+
\xi_1^+ + c_- \xi_1^-\rangle = | c_+ |^2$$ and on the
opposite vertex with the probability  $p_3 = | c_- |^2$. So
the numbers $~p_j,p_k$ must be interpreted due to quantum
mechanics as the squares of moduli of {\it probability
amplitudes}. The identities obtained by us earlier for the
classical game make it possible to compare each of the four
pairs of numbers $\{p_1, p_3\}$, $\{p_2, p_4\}$, $\{q_1,
q_3\}$, $\{q_2, q_4\}$ {\it separately} with the
corresponding {\it conditional} probabilities. Compare
formula(12)for the quantum average
$$E_{\varphi\otimes\psi}\widehat{{\cal H}}_A =(ap_1 q_3
+cp_3 q_1) + (bp_2 q_4 + dp_4 q_2) $$ with
formula~(\ref{classic}) used for calculation of the
classical average by the use of conditional probabilities
and conditional expectation values:
 \begin{equation}\label{fpw}
E{\cal H}_A =(ap_{13}^1 q_{13}^3 + cp_{13}^3 q_{13}^1)\cdot
P_{13} + (bp_{24}^2 q_{24}^4 + dp_{24}^4 q_{24}^2)\cdot
P_{24}
\end{equation}
In the case when the corresponding pairs of the squares of
moduli of amplitudes are equal to classical conditional
probabilities $$p_1=p_{13}^1,\quad p_3=p_{13}^3,\quad
p_2=p_{24}^2,\quad p_4=p_{24}^4 $$ $$q_1=q_{13}^1,\quad
q_3=q_{13}^3,\quad q_2=q_{24}^2,\quad q_4=q_{24}^4 $$ one
obtains an interesting result: the "wise" Alice gets a
larger payoff than that which is obtained by her "foolish"
copy: $$E_{\varphi\otimes\psi}\widehat{{\cal H}}_A
\geqslant E{\cal H}_A$$ Notice, however, that to have this
one must have the situation where the squares of moduli of
the probability amplitudes in "quantum" Nash equilibrium
are equal to the conditional probabilities obtained by
applying the apparatus of the classical game theory.
However there is no foundation for such an equality. In
fact, if the equilibrium point of the classical game is
searched on the set of nonnegative numbers with constrains
$$x_1+x_2+x_3+x_4=1,\qquad y_1+y_2+y_3+y_4=1$$ then in the
quantum game case for the squares of moduli of {\it the
probability amplitudes} besides {\it the explicit} linear
relations $$p_1+p_3=1,\qquad p_2+p_4=1,\qquad
q_1+q_3=1,\qquad q_2+q_4=1$$ there are implicit relations
due to unitary dependence~(\ref{Hquant}) of projection
operators with even and odd indices. The difference between
the two pictures is due to the fact that in the quantum
situation the formula of full probability~(\ref{fpw}) is
broken: the average payoff to Alice is equal to {\it the
sum} of conditional probabilities and not to their {\it
mixture}: $~P_{13}=P_{24}=1$. This again demonstrates that
it is impossible to find a quantum equilibrium point by use
of only formula~(\ref{amplitude}) of the average payoff as
a function of eight variables $\{p_j, q_k\}$ with the
relations written before between them. The equilibrium is
defined not by the combination of {\it the squares of
moduli} of the amplitudes of Alice and Bob but by the
combination of {\it wave functions} $\varphi\in{\sf H}_A$
and $\psi\in{\sf H}_B$.

%%%%%%%%%%%%%%%%%%%%%%%%%%%%%%%%%%%%%%%%%%%%%%%%%%%%%%
\bigskip
\section{\point Behaviour of the player as realisation of logic}

Besides the amplitudes, characterising the behaviour of
players, our model has two additional structural
characteristics: unitary rotations $~u~$ and $~v~$, giving
an operator representation of the ortholattices of Alice
and Bob up to unitary equivalence. Each of these operators
can be characterised by the angle between eigensubspaces of
even and odd order. Denote by $\theta_A$  the angle between
direct lines corresponding to the largest eigenvectors of
operators  $\widehat{\alpha}_1$ , $\widehat{\alpha}_2$ of
Alice and $\theta_B$ the analogous angle characterising the
realisation of the logic of Bob. Differently from the
amplitudes characterising the behaviour of the players,
being the {\it variables} of the model, the angles are its
{\it parameters}.These parameters characterise the {\it
type} of player, making it possible to consider logic as
some factor forming the  behaviour. Some sense of these
values can be the following: the angles characterise the
connections between choices of the diagonals of the square
or some preferences for this or that adjacent vertex. The
angles $\theta_A,~\theta_B$   define commutation relations
for corresponding operators. Consider for example the
operators $\widehat{\alpha}_1$, ¨ $\widehat{\alpha}_2$.
Taking as an orthonormal basis in the space $~H_A~$of Alice
the eigenbasis of the operator $\widehat{\alpha}_1$ let us
write the matrices of this pair of operators
$$\widehat{\alpha}_1=\rm \left(
\begin{array}{rr}
1 & 0   \\ 0 & 0   \\
\end{array}
\right), \hspace{30pt} \widehat{\alpha}_2 = \rm \left(
\begin{array}{lr}
\cos^2{\theta}_A  & \sin{\theta}_A \cos{\theta}_A  \\
\sin{\theta}_A \cos{\theta}_A  &  sin^2{\theta}_A
\\
\end{array}
\right)$$ where $\theta_A$ is the angle on which eigenbasis
of the operator $\widehat{\alpha}_2$ is rotated relative to
the eigenbasis of the operator $\widehat{\alpha}_1$.
Calculating the commutator of these matrices one obtains:
$$[\widehat{\alpha}_1,\,\widehat{\alpha}_2]
=\frac{i}{2}\cdot\sin 2\theta_A\rm \left(
\begin{array}{rr}
0 & -\,i   \\ i & 0   \\
\end{array}
\right)$$ Analogous commutation relations are obtained for
Bob's operators. Non-commutativity of the operators of the
logic representation expresses evident, without any
mathematics, the dependence of the results of the  game on
the {\it order} of acts. So Bob's state is different
depending on the order: "1", then "2" or "2", then "1", in
which he received Alice's questions . So in the quantum
model one takes into account not only the interests of the
players represented in the payoff matrix but in some sense
their personal features on which depends the level of
realisation of interests.

\bigskip
\section{\point In search of the quantum equilibrium.}

The definition of the Nash equilibrium for the quantum case
is not much different from the classical case~(\ref{NE})
and can be written as $$E\widehat{{\cal H}}_A
(\varphi^0,\psi^0)\geqslant E\widehat{{\cal H}}_A
(\varphi,\psi^0),\qquad E\widehat{{\cal H}}_B
(\varphi^0,\psi^0)\geqslant E\widehat{{\cal H}}_B
(\varphi^0,\psi)$$ It is convenient to find the equilibrium
points in the coordinate form. To do this let us fix in the
space of strategies of Alice $~H_A~$ eigenbases $\{\xi_1^+,
\xi_1^-\}$ ¨ $ \{\xi_2^+, \xi_2^-\}$ corresponding to two
projectors $\widehat{\alpha}_1, \widehat{\alpha}_2$ and let
us do the same for Bob, taking bases $\{\eta_1^+,
\eta_1^-\}$ ¨ $ \{\eta_2^+, \eta_2^-\}$. The angles between
the largest eigenvectors denote as $\theta_A$ and
$\theta_B$. Then one can write in the quantum payoff
function $$E\widehat{{\cal H}}_A({\varphi, \psi}) =ap_1 q_3
+cp_3 q_1 + bp_2 q_4 + dp_4 q_2 $$ the squares of moduli of
the amplitudes $~p_j,p_k~$ as $$ p_1=\cos^2\alpha,\quad
p_3=\sin^2\alpha,\quad p_2=\cos^2(\alpha-\theta_A),\quad
p_4=\sin^2(\alpha-\theta_A), $$
    $$q_1=\cos^2\beta,\quad q_3=\sin^2\beta,\quad
    q_2=\cos^2(\beta-\theta_B),\quad q_4=\sin^2(\beta-\theta_B),$$
where $\alpha, \beta$  are the angles of vectors $\varphi,
\psi$ to the corresponding axises. For values of angles one
can take the interval $[0^0; 180^0]$. In the result the
problem of search of the equilibrium points of the quantum
game became the problem of finding a minimax of the
function of two angle variables $$F(\alpha, \beta)=
a\cos^2\alpha \sin^2\beta +c\sin^2\alpha \cos^2\beta +$$
$$+b\cos^2(\alpha-\theta_A) \sin^2(\beta-\theta_B) +
d\sin^2(\alpha-\theta_A) \cos^2(\beta-\theta_B)$$ on the
square $[0^0; 180^0]\times[0^0; 180^0]$. In other words our
quantum game occurred to be {\it an infinite antagonistic
game} of two persons on the square. Solving such games in
pure strategies, i.e. search for saddle points of the
function $F(\alpha, \beta)$  is a difficult problem. It is
known~\cite{Worobyev} that for the existence of the saddle
points properties of continuity and smoothness of the
payoff function are not enough. Present theorems of
existence use properties of convexity of this function.
However in our case we don't have these properties. One can
find examples of some values of the elements $a,b,c,d$ of
the payoff matrix~(\ref{h2}) when the function $F(\alpha,
\beta)$  doesn't have saddle points. Not much better is the
situation with methods of search of the saddle points.
Differently from the geometrical saddle points the
conditions of the Nash equilibrium are not just putting to
zero values of the corresponding partial derivatives. So in
the situation of absence of simple analytical solutions one
must look for numerical methods. To do calculations we use
an algorithm based on the construction of "curves of
reaction" or "curves of the best answers" of the
participants of the game.

The definition of  curves of reaction is based on the
following consideration. If Alice knew what decision Bob
will take  she could make an {\it optimal} choice. But the
essence of the game situation is that she doesn't know it.
She must take into account his different strategies and on
each possible act of the partner she must find the optimal
way to act. Her considerations look like considerations of
the player, expressed by the formula: "if he does this,
then I shall do that". Bob thinks the same way. So one must
consider two functions $\alpha = {\cal R}_A (\beta)$ and
$\beta = {\cal R}_B (\alpha)$ the plots of which are called
the curves of reactions of Alice and Bob. Due to the
definition of these functions $$\max_{\alpha} F(\alpha,\,
\beta)= F({\cal R}_A (\beta),\, \beta),\qquad \min_{\beta}
F(\alpha,\, \beta)=F(\alpha,\, {\cal R}_B (\alpha))$$ It is
easy to see that intersections of curves of reaction give
points of Nash equilibrium. Numerical experiments show that
dependent on the values of the parameters $a,b,c,d$ of the
payoff function and the angles characterising the type of
player one has qualitatively different pictures.
Intersections can be absent, there can be one intersection
and lastly there can be the case with two equilibrium
points with different values of the payoff of the game,
which is absent in the case of the classical matrix game.

\section{\point Examples}

{\bf 1. \em Two equilibrium} points arise in the case of
the payoff matrix:

\begin{table}[h]\label{h2}
\begin{center}
\begin{tabular}{|c||c|c|c|c|}
\hline $A \backslash B$ &   {\bf 1} & {\bf 2} & {\bf 3} &
{\bf 4}
\\ \hline\hline
 {\bf ~1~} & ~0~ &  ~0~ & ~3~ & ~0~ \\ \hline
 {\bf ~2~} & ~0~ &  ~0~ & ~0~ & ~3~ \\ \hline
 {\bf ~3~} & ~5~ &  ~0~ & ~0~ & ~0~ \\ \hline
 {\bf ~4~} & ~0~ &  ~1~ & ~0~ & ~0~ \\ \hline
\end{tabular}
\end{center}
\end{table}\vspace{-15pt}
\noindent and an operator representation of the
ortholattice corresponding to  angles $\theta_A=10^0$,
$\theta_B=70^0$. One of the equilibrium points is  inside
the square, the other one is on it's boundary
(see~Fig.~\ref{fig5}).

%%%%%%%%%%%%%%%%%%%%%%%%%%%%%%%%%%%%%%%%%%%%%%%%%%%%%
\begin{figure}[ht]
\begin{center}
\includegraphics[height=0.45\textwidth]{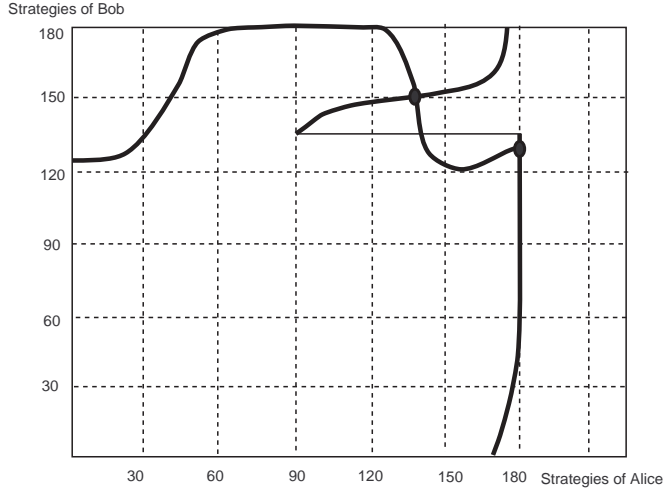}
\end{center}
\vspace{-15pt} \caption{\em Two points of Nash equilibrium}
\label{fig5}
\end{figure}
%%%%%%%%%%%%%%%%%%%%%%%%%%%%%%%%%%%%%%%%%%%%%%%%%%%%%

\noindent The curves of reaction in this case happen to be
{\it discontinuous}. For convenience the discontinuities
are shown by thin lines. The discontinuous character of the
curve of reaction of Alice made it impossible for one more
equilibrium point to occur. One of the equilibrium takes
place for $\alpha=145,5^0$, $\beta=149,5^0$ and gives the
following values for the squares of moduli of amplitudes:

for Alice $p_1=0,679; p_2=0,509; p_3=0,321; p_4=0,491;$

for Bob $q_1=0,258;q_2=0,967;q_3=0,742;q_4=0,033.$

\noindent The price of the quantum game, i.e. the
equilibrium value of the profit for Alice in this case is
equal to $E\widehat{\cal H}_A=2.452$. The second
equilibrium point corresponds to angles $\alpha=180^0$,
$\beta=123,5^0$ and the squares of the amplitude moduli

for Alice $p_1=1.000; p_2=0,967; p_3=0.000; p_4=0.033;$

for Bob $q_1=0,695;q_2=0.646;q_3=0.305;q_4=0.354.$

\noindent The price of the game in the second equilibrium
point is equal to $E\widehat{\cal H}_A=1.926$. For the {\it
classical} game with the same payoff matrix (see section~4)
the price of the game happens to be smaller and is equal to
{\large ${15\over 28}$}~=~$0.536.$ Differently from the
quantum game the classical game has only{\it one}
equilibrium point which is obtained for the following
frequencies

for Alice {\large $x_1={5\over 28}$; $x_2={5\over 28}$;
$x_3={3\over 28}$; $x_4={15\over 28};$}

for Bob {\large $y_1={3\over 28}$; $y_2={15\over 28}$;
$y_3={5\over 28}$; $y_4={5\over 28}$.}

\noindent To compare the quantum game with the classical
one the conditional probabilities for the choice of
vertices of the square after the choice of the diagonal are
found to be:

for Alice: {\large $p^1_{13}={5\over 8}$; $p^3_{13}={3\over
8}$; $p^2_{24}={1\over 4}$; $p^4_{24}={3\over 4}$;}

for Bob : {\large $q^1_{13}={3\over 8}$; $q^3_{13}={5\over
8}$; $q^2_{24}={3\over 4}$; $q^4_{24}={1\over 4}$;}

\medskip

\noindent and the conditional average
payoffs~(\ref{Hal13},\ref{Hal24}) for each diagonal are:
$$~E_{13}{\cal H}_A=1.875; \qquad E_{24}{\cal H}_A=0.75.$$

\noindent The price of the classical game is obtained by
multiplication of these expressions on the probabilities of
the corresponding conditions given in section(4). Terms of
the quantum payoff, associated with these conditional
averages $E_{\varphi\otimes\psi}\widehat{{\cal H}}_A$ in
case of the first equilibrium point are
\begin{center}$~ap_1q_3 + cp_3q_1=1.927~;$ \quad $~bp_2q_4 +
dp_4q_2=0.525~$ \end{center}

\noindent For the second equilibrium point one has:
\begin{center} $ap_1q_3 + cp_3q_1=0.915$ ;\quad $bp_2q_4 +
 dp_4q_2=1.048$.\end{center}

\noindent {\bf 2. \em A unique equilibrium} is observed for
example in the case when all nonzero payoffs are equal and
are equal to one and for equal angles $\theta_A=45^0$,
$\theta_B=45^0$. The equilibrium point is located in the
upper right vertex of the square (see Fig.~\ref{fig6}):

%%%%%%%%%%%%%%%%%%%%%%%%%%%%%%%%%%%%%%%%%%%%%%%%%%%%%
\begin{figure}[ht]
\begin{center}
\includegraphics[height=0.45\textwidth]{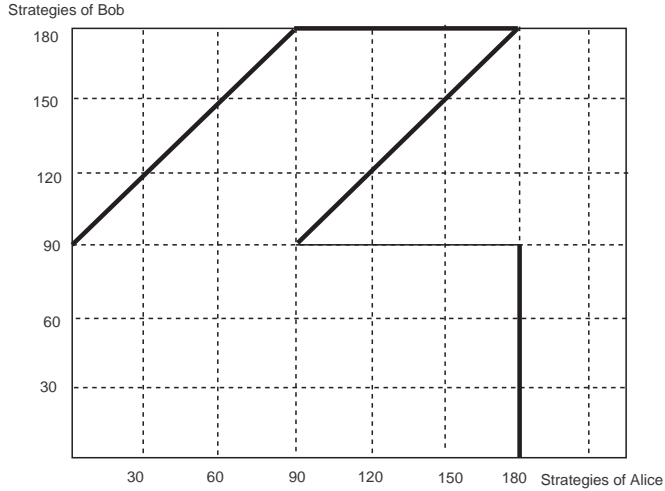}
\end{center}
\vspace{-15pt} \caption{\em The unique Nash equilibrium}
\label{fig6}
\end{figure}
%%%%%%%%%%%%%%%%%%%%%%%%%%%%%%%%%%%%%%%%%%%%%%%%%%%%%

\noindent The curve of Bob's reaction is shown on the
Fig.~\ref{fig6} as {\it continuous} while the analogous
curve of Alice is discontinuous when Bob is using the
strategy corresponding to the angle $\beta=90^0$.  To make
it more explicit the discontinuity is shown by drawing the
thin line. In reality {\it both} lines are discontinuous.
This becomes evident if one prolongs both functions on the
whole real axis taking into account the periodicity: the
plots of one of them is obtained by the shift of the other
one on the halfperiod-$90^0$. The squares of the amplitude
moduli in this case have the following values

for Alice: $~p_1=1~$; $~p_2=0.5~$; $~p_3=0~$; $~p_4=0.5~$;

for Bob: $~q_1=1~$; $~q_2=0.5~$; $~q_3=0~$; $~q_4=0.5~$.

\noindent The payoff of the "wise" Alice in this case is
$~E\widehat {\cal H}_A=0.5~$ while her classical copy gets
only $0.125$. Differently from the quantum case for the
classical players with such a payoff function all vertices
of the square are equally probable.

The unique equilibrium located {\it inside} the square
takes place for the initial payoff matrix $a~=~3,\;
b~=~3,\; c~=~5,\; d~=~1$ and angles $\theta_A=15^0$,
$\theta_B=35^0$ (see Fig.~\ref{fig7}).

%%%%%%%%%%%%%%%%%%%%%%%%%%%%%%%%%%%%%%%%%%%%%%%%%%%%%
\begin{figure}[ht]
\begin{center}
\includegraphics[height=0.45\textwidth]{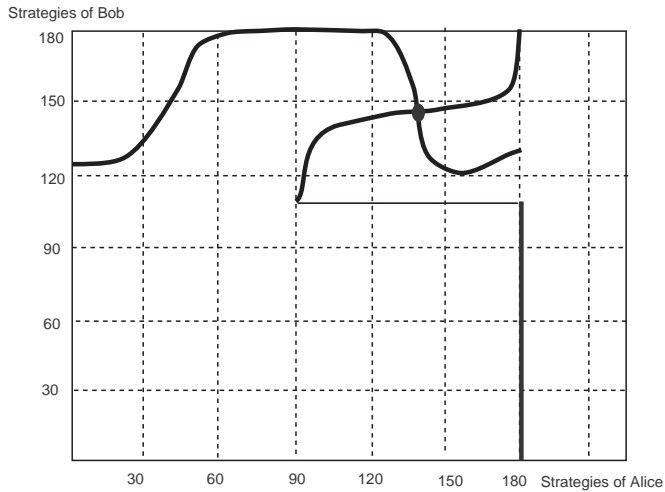}
\end{center}
\vspace{-15pt} \caption{\em The other example of the unique
Nash equilibrium} \label{fig7}
\end{figure}
%%%%%%%%%%%%%%%%%%%%%%%%%%%%%%%%%%%%%%%%%%%%%%%%%%%%%

\noindent {\bf 3. \em Absence of equilibrium} is perhaps
one of the most interesting phenomena, because as it is
known for classical matrix games, equilibrium in mixed
strategies always exist. One can obtain absence of
equilibrium by taking the same payoff matrix for which one
as well as two points of equilibrium were found. For this
it is sufficient to take the operator representation of the
ortholattice with typical angles: $\theta_A=30^0$,
$\theta_B=20^0$ (see Fig.~\ref{fig8}).

%%%%%%%%%%%%%%%%%%%%%%%%%%%%%%%%%%%%%%%%%%%%%%%%%%%%%
\begin{figure}[ht]
\begin{center}
\includegraphics[height=0.45\textwidth]{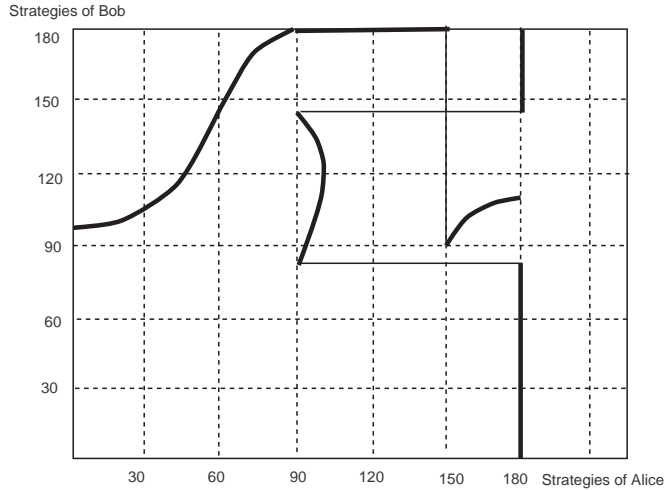}
\end{center}
\vspace{-15pt} \caption{\em Absence of Nash equilibrium}
\label{fig8}
\end{figure}
%%%%%%%%%%%%%%%%%%%%%%%%%%%%%%%%%%%%%%%%%%%%%%%%%%%%%

Absence of equilibrium in this case as it is seen from the
Fig.~\ref{fig8} is due to the discontinuity of the
functions of reaction which is impossible in the classical
case. We met this phenomenon in the first example when two
equilibrium points were obtained. This last example shows
the importance of the {\it realisation} of a
non-distributive lattice. In the language of the game
theory one can understand it as follows: having the same
interests the players can form their behaviour
qualitatively in different ways. So the mathematician can
give to the client, for example to  Alice, strategic
recommendations: how she can organise the style of her
behaviour to make the profit larger for the same payoff
conditions. For this, however, he must know the choice of
the representation of Bob's logic.

\bigskip
\section*{Concluding remarks}
The construction and analysis of our models show that the
main difference between classical and quantum points of
view on observable phenomena is expressed in the {\it way
of calculation of averages}. In fact, if one remembers the
first work of M.Planck on the spectra of radiation of a
perfect black body one will see that the correct formula
was obtained on the basis of a postulate, leading to the
other than classical way of calculating oscillator average
energy. Instead of the uniform distribution for degrees of
freedom one used an averaging based on totally different
statistics. The subsequent history of quantum physics is in
some sense the history of the development of the concept of
the average: important characteristics of the objects of
the microworld are manifested in their "statistics". The
development of the mathematical formalism of quantum
mechanics is also strongly based on the same idea. Positive
functionals on non-commutative algebras with involution in
GNS construction are nothing but averages. The difference
between the quantum and classical concepts of average
consists in the fact that  there are questions in quantum
mechanics that cannot have simultaneous answers, i.e. the
ortholattice of answers is not a Boolean algebra, the
distributivity rule is broken in it. Practically all modern
theories of quantum mechanics more or less explicitly
suppose that the corresponding lattice is the lattice of
closed subspaces of Hilbert space. This supposition has a
constructive character and leads to operator
representations and it is just this scheme of calculating
the averages that was realised in this work. In what sense
and what is the justification of using the apparatus of
quantum mechanics in game theory when one deals with
properties of {\it macroscopic} objects? The woman taking
in her hand the diamond and feeling its anomalous heat
capacity can know nothing about the formula of
Einstein-Debye and about the specific procedure of
averaging, explaining the observed phenomenon. However, one
who knows it will not be surprised by the attempts to
explain some features of macrophenomena by the specific way
of calculating the averages. The exact answer to our
question implies an analysis of the logical structure of
the investigated phenomena. If the logic adequately
representing the experience is non-distributive then
classical procedures of calculating the averages lead, as
we saw, to contradictions and one must use another
apparatus. If the obtained lattice happens to be the
lattice of subspaces, then the answer is given by the
Gleason's theorem~\cite{Gleason} saying that probability
measures on the lattice of projectors have strictly
definite form. So if one is solving the problem of
averaging of the payoff function, taking into account {\it
logical conditions} of the players, then going to
quantization in some cases is predestined.

One can only be surprised that the structure of the
classical expression of the payoff function is such as if
it were specially invented to put there instead of the
two-valued function the operators. This leads to the
general scheme of quantization of games. In any case if one
considers not antagonistic but {\it bimatrix} game when the
interests of the players are not strictly antagonistic, the
scheme of quantization will be the same. This problem
however as well as games of many persons goes beyond the
contents of this paper. The problem of non-uniqueness of
orthocomplements in the lattices is also goes beyond our
paper.It exists even in case of our lattice: for negations
one can take correspondence between "even" and "odd"
questions. In the general case this problem, touching the
theory of representations of internal symmetries of
non-classical logics can lead also to some natural
application of the formalism of quantum mechanics. The
important point in our scheme is taking care in the
difference between the logic and its operator
representation. One sees that there exists a continuum of
classes of unitary equivalent representations,
parameterised by the angles of the type~$\theta_A$,
$\theta_B$.  Each of these representations can be described
by the commutation relations for the corresponding
non-commuting operators. In the players logic these pairs
of projectors correspond to mutually complementary
questions, neither of which is the negation of the other.
As it was shown previously non-commutativity is connected
with the dynamics of the game -- the order in which
questions are posited. Considering the quantum game we had
the problem of {\it existence} of equilibrium. As was seen
from the examples their existence or absence is connected
not only with the structure of the payoff matrix but also
with the representation of the lattice of properties.
Absence of equilibrium is also connected with the
representation of the behaviour of the players by pure
states-by {\it vectors} in Hilbert spaces, when the game
situation is represented by some {\it resolved} element of
their tensor product. The notion of equilibrium can be
enlarged as it is done in the classical game theory taking
mixed enlargement on the basis of the density matrices.
Here we didn't consider {\it entangled} states. As was
shown~\cite{GrRZ2} in the case of entangled states one must
deal with more complex non-distributive lattices. The
search of equilibrium among entangled non-factorisable
states can lead to success in proving the existence of
equilibrium in the general case of quantum games.

%\medskip
\section*{\normalsize Acknowledgements}
One of the authors (A.A.G.) is indebted to the Foundation
of the Ministry of Education of Russia, grant E0-00-14 for
the financial support of this work.
%%%%%%%%%%%%%%%%%%%%%%%%%%%%%%%%%%%%%%%%%%%%%%%%%%%%%%%

%\medskip


\begin{thebibliography}{99}

\bibitem{GrRZ1}{\em Grib A.A., Zapatrin R.R.}
Int. Journ. Th. Phys. {\bf  29} (2),  (1990), p.113.

\bibitem{GrRZ2}{\em Grib A.A., Zapatrin R.R.}
Int. Journ. Theor. Phys. {\bf  30} (7),  (1991), p.949.

\bibitem{Gribook}{\em Grib A.A.} Breaking  of Bell's
inequalities  and the problem of measurement in the quantum
theory. Lectures for young scientists  -- JINR, Dubna,
(1992), (in Russian)


\bibitem{Waldir}{\em Grib A.A., Rodrigues Jr.W.A.}
Nonlocality in Quantum Physics // Kluwer Academic / Plenum
Publishers -- N.Y., Boston, Dodrecht, London, Moscow,
(1999).

\bibitem{Deutsch}{\em Deutsch D.} Quantum
Theory of Probability and Decisions. --  Proc. R. Soc.
Lond. Acad., (1999). (quant-ph/9906015)

\bibitem{Finkelstein}{\em Finkelstein J.} Quantum
Probability from Decisions Theory? -- SJSU-99-20, June
(1999).

\bibitem{Polley}{\em Polley I.} Reducing the Statistical
Axiom. -- Phys. Dep. Oldenburg Univ. D-26111, July (1999).

\bibitem{Eisert}{\em Eisert J. et all} Phys. Rev. Lett.
{\bf 83}, (1999), p.3077.

\bibitem{Ekert}{\em Ekert A.K.} Phys. Rev. Lett. {\bf 67},
(1999), p.661.

\bibitem{Marinatto}{\em Marinatto L., Weber T.} Phys.
Lett. A {\bf 272}, (2000), p.291.

\bibitem{Moulin}{\em Moulin H.} Th\'eorie des jeux
pour l'\'economie et la politique. -- Hermann, Paris,
(1981).


\bibitem{Neumann}{\em von Neumann J., Morgenstern O.}
Theory of games and economic behaviour. -- Princeton.
Princeton Univ. Press, (1953).

\bibitem{Kemeny}{\em Kemeny J.D., Thompson G.L.} The effect
of psychological attitudes on the outcomes of games //
Contributions to theory of games -- {\bf 3}, Princeton,
(1957) pp.273-298.


\bibitem{Birkhoff}{\em Birkhoff G.} Lattice Theory --
AMS Colloc. Publ., vol. 25, AMS, Providence, Rhode, Island,
(1993).

\bibitem{Gratzer}{\em Gr\"atzer G.} Lattice Theory --
San Francisco: Freeman \& Co., (1971).

\bibitem{Worobyev}{\em  Worob'ev N.N.} Foundations of Game
Theory. Non-cooperative games -- Birkh\"auser Verlag.
Basel, Boston, Berlin, (1994).

\bibitem{Gleason}{\em Mackey G.W.} The Mathematical Foundations
of Quantum Mechanics -- W.A.Benjamin, INC. N.Y., Amsterdam,
(1963).

\bibitem{Sudbery} {\em Sudbery A. } Quantum Mechanics
and particles of nature -- Cambridge. Univ. Press., (1986).

\bibitem{Parfionov}{\em Parfionov G.N., Zapatrin R.R.}
Linear programming method for finding orthocomplements in
finite lattices -- Int. Journ. of  Th. Phys, {\bf 37},
(1998) pp.211-213.
\end{thebibliography}
\end{document}